\documentclass[prb,showpacs,preprintnumbers,amsfonts,amssymb,amsmath,floats,twocolumn,aps]{revtex4}

\usepackage{graphicx}
\usepackage{dcolumn}
\usepackage{bm}
\usepackage[final,dvips]{epsfig}
\usepackage{bm}
\usepackage{color}

\newcommand{\refeq}[1]{Eq.\ (\ref{eq:#1})}
\newcommand{\labeq}[1]{\label{eq:#1}}

\newcommand{\ff}[1]{{\boldsymbol #1}}
\newcommand{\ca}[1]{{\cal #1}}

\begin{document} 
  
\title{Macrospin approximation and quantum effects in models for magnetization reversal}

\author{Mohammad Sayad, Daniel G\"utersloh and Michael Potthoff}

\affiliation{I. Institut f\"ur Theoretische Physik, Universit\"at Hamburg, Jungiusstra\ss{}e 9, 20355 Hamburg, Germany}

\begin{abstract}
The thermal activation of magnetization reversal in magnetic nanoparticles is controlled by the anisotropy-energy barrier. 
Using perturbation theory, exact diagonalization and stability analysis of the ferromagnetic spin-$s$ Heisenberg model with coupling or single-site anisotropy, we study the effects of quantum fluctuations on the height of the energy barrier. 
Opposed to the classical case, there is no critical anisotropy strength discriminating between reversal via coherent rotation and via nucleation/domain-wall propagation.
Quantum fluctuations are seen to lower the barrier depending on the anisotropy strength, dimensionality and system size and shape.
In the weak-anisotropy limit, a macrospin model is shown to emerge as the effective low-energy theory where the microscopic spins are tightly aligned due to the ferromagnetic exchange.
The calculation provides explicit expressions for the anisotropy parameter of the effective macrospin.
We find a reduction of the anisotropy-energy barrier as compared to the classical high spin-$s$ limit. 
\end{abstract} 
 


\maketitle 

\section{Introduction}

Nanosystems of interacting magnetic moments have attracted much interest due to ongoing technological advances in the field of magnetic data-storage devices.
With decreasing system size, ferromagnetic particles on solid surfaces \cite{Wie09} or ferromagnetic molecular magnets like Fe$_8$ and Mn$_{12}$, \cite{Sch04} for example, are found in a single-domain state. 
The question of the stability of the ferromagnetic state against different kinds of thermal and quantum fluctuations is a crucial technological issue which at the same time provides serious challenges to a theoretical modeling and understanding.

The magnetic properties of a ferromagnetic nanoparticle can be described by a Heisenberg model
\begin{equation}
  H_J = - \frac{1}{2} \sum_{ij} J_{ij} \ff s_i \ff s_j
\labeq{ham}
\end{equation}
with positive exchange coupling $J_{ij}>0$ between microscopic spins $\ff s_i$ and $\ff s_j$.
The spins give rise to magnetic moments which are assumed to be localized at the sites $i=1,...,L$.
The latter may constitute a $d$-dimensional lattice of finite size with $L$ sites in total and constant exchange $J=J_{ij}$ between nearest neighbors $i$ and $j$ only.
In a ground state of the model all spins are perfectly aligned, and the particle is ferromagnetic.
However, this state is unstable against thermal fluctuations.
Strictly speaking, for a system of finite size, the ferromagnetic state is destroyed at any finite temperature by fluctuations originating from a coupling of the system to an external heat bath.

On the other hand, anisotropic contributions to the Hamiltonian, such as a single-site anisotropy of the form
\begin{equation}
  H_D = - D \sum_{i} s^2_{iz} \; ,
\labeq{hamani1}
\end{equation}
which for $D>0$ distinguishes the $z$-axis as the easy axis, stabilize the ferromagnetic state. 
Inclusion of $H_D$ leads to a model with only two degenerate ground states in which all spins are pointing into the $+z$ or $-z$ direction, respectively. 
The anisotropic ferromagnetic Heisenberg model $H=H_J+H_D$ in fact represents the simplest model to study magnetization reversal, i.e.\ the thermally activated transition between two ground states across an energy barrier induced by the anisotropy. \cite{Wer01}
The rate of magnetization switching and thus the magnetic stability obviously crucially depends on the height of the anisotropy-energy barrier $\Delta E$.

\begin{figure}[b]
  \includegraphics[width=0.95\columnwidth]{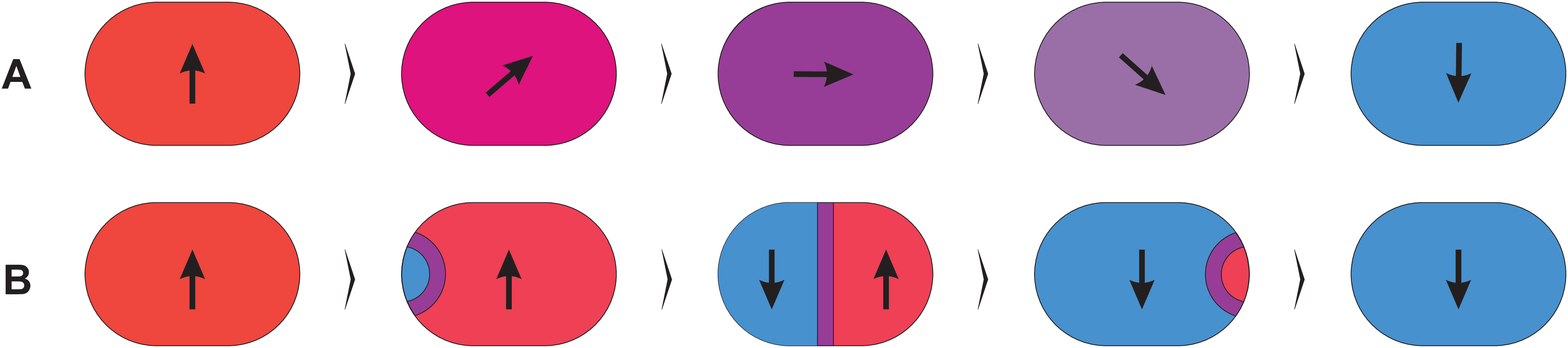}
\caption{
Two different models for magnetization reversal. A: Coherent rotation. B: Nucleation and domain-wall propagation.
Figure inspired by Ref.\ \onlinecite{KHS+09}. 
}
\label{mechs}
\end{figure}

For weak anisotropy $D \ll J$, it is self-evident to consider the spins as tightly coupled by the exchange interaction and forming a huge macrospin,
\begin{equation}
   \ff S = \sum_i \ff s_i \: ,
\labeq{macrospin}
\end{equation}
the dynamics of which can be approximated by an effective model
\begin{equation}
  H_{\rm macro} = - D_{\rm macro} S^2_{z} + \mbox{const.} \: .
\labeq{hammacro}
\end{equation}
with an anisotropy barrier
\begin{equation}
  \Delta E = D_{\rm macro} S^2 \: .
\end{equation}
In this macrospin model, magnetization reversal takes place at zero temperature by suppression of the barrier due to an external magnetic field as described by the Stoner-Wohlfarth model. \cite{SW48}
At finite temperature $T$, reversal may be caused by thermal activation. 
For low $T$, the switching frequency is exponentially small and a ferromagnetic state is stable over macroscopically relevant times while for high $T$, in the so-called superparamagnetic state, the magnetization vanishes on the time scale of the measurement.
Classical theory \cite{Nee49,Bro63} for superparamagnetic dynamics based on the Landau-Lifshitz equation \cite{LL35} and including a stochastic Langevin field to simulate a thermal bath leads to an Arrhenius law for the thermal activation rate $\Gamma \propto \exp(- \beta \Delta E)$.
This Neel-Brown law again emphasizes the role of the energy barrier $\Delta E$ for the magnetic stability of the particle.

For stronger anisotropy, the macrospin approximation is no longer applicable as magnetization reversal preferably takes place by nucleation and domain-wall propagation (see Fig.\ \ref{mechs}).
Here, classical many-spin models \cite{KG05} are frequently used to study the reversal mechanism and in particular the transition from coherent rotation to nucleation and domain-wall propagation with increasing strength of the anisotropy. 
In particular, classical Monte-Carlo simulations are employed \cite{HN98b} where Monte-Carlo update steps are related to physical time propagation. 
As a function of the model parameters, reversal mechanisms different from coherent rotation, can be identified and ``phase boundaries'' can be found in this way.
The strong influence of the shape of the nanoparticles on the switching rate, as observed experimentally, \cite{BPKW04,KHS+09} can be explained by theoretical analysis of domain-wall propagation. \cite{Bra93,Bra94}

The macrospin has a high spin quantum number $S$ and may be treated to a good approximation classically.
There are, however, different types of quantum effects to be considered:
For example, quantum tunneling as a reversal mechanism is competing with thermally induced reversal. \cite{ADiVS92,WS99,DeRHD+02}
In case of not too small systems, tunneling is significant only for systems extremely isolated from any dissipative environment at ultralow temperatures.
For a macrospin coupled to a conduction band, Kondo screening is another issue.
Since $S>1/2$ there is an underscreened Kondo effect, the development of which is hampered by the anisotropy as has been shown by time-dependent numerical renormalization group. \cite{RWH08}
As the quantum character of the macrospin is not accounted for by simple Langevin dynamics, \cite{Bro63} different approaches \cite{ZGP06,KCT08,BL05} have been suggested to treat the macro- or many-spin dynamics in contact with bosonic baths by means of quantum master-equation approaches and to determine, e.g., the blocking temperature.
The importance of a many-spin model, for example, is demonstrated by studies of the probability density function of the macrospin obtained by replacing thermal with Markov processes. \cite{BL05}
Differences between the classical-spin limit and large quantum spins have been discussed. \cite{Gar08}
 
Here we reconsider the question of competing reversal mechanisms by taking a quantum many-spin model as the starting point.
The competition between coherent rotation and nucleation can be addressed in a less ambitious way by studying the {\em static} properties of an anisotropic quantum Heisenberg model $H = H_J +H_{\rm ani}$.
As the single-site anisotropy (\ref{eq:hamani1}), which is frequently studied in the classical context, cannot be used for the lowest spin $s=1/2$ in a quantum model since trivially $\ff s_i^2=\mbox{const}$, we also consider a coupling anisotropy of the form
\begin{equation}
  H_\Delta = - \frac{1}{2} \sum_{ij} \Delta_{ij} s_{iz} s_{jz} \: ,
\labeq{hamani2}
\end{equation}
i.e.\ $H_{\rm ani} = H_D$ or $H_{\rm ani} = H_\Delta$.
This model will be investigated by different complementary techniques including perturbation theory in $H_{\rm ani}$, exact diagonalization and the Lanczos approach for systems with small size $L$ and classical stability analysis in the high-spin limit.

There are different goals of the present paper:
As the status of the macrospin approximation is less clear in the quantum case, we aim at a strict derivation of the macrospin model in the weak-anisotropy limit. 
In this way it should be possible to relate the effective anisotropy strength of the macrospin to the parameters of the underlying many-spin model.
The dependence on the spin quantum number $s$ should exhibit quantum effects for small $s$ and recover the classical result for $s \to \infty$.
Furthermore, beyond the weak-anisotropy limit, the breakdown of the macrospin approximation and the competition between the different reversal mechanisms, depending on system size, coordination, dimensionality and on $s$, will be investigated.
Our goal is to understand whether or not it makes a qualitative difference for the transition from coherent rotation to nucleation with increasing anisotropy strength if starting from a classical or from a quantum spin model.
Finally, the high-spin limit is addressed as a reference and for comparison with the  quantum-mechanical calculations.

The paper is organized as follows:
In the next section \ref{sec:ed} we briefly introduce the model, the basic concepts and notations by referring to exact-diagonalization results. 
Sec.\ \ref{sec:pert} present the results of first-order perturbation theory for weak anisotropy and the derivation of the macrospin model.
Its limitations are discussed in Sec.\ \ref{sec:lim} on the basis of calculations employing the Lanczos technique.
The dependence of the anisotropy-energy barrier on the spin-quantum number $s$ and classical stability analysis in the high-spin limit is addressed in Sec.\ \ref{sec:class}.
Finally, Sec.\ \ref{sec:con} concludes the paper.

\section{Exact-diagonalization results for the anisotropy barrier}
\label{sec:ed}

Due to the SU(2) spin-rotation symmetry of the Heisenberg model, \refeq{ham}, the total (or macro) spin, \refeq{macrospin}, is a conserved quantity: $[\ff S, H]=0$.
We consider the eigenstates of $H_J$:
\begin{equation}
  H_J | n, S, M \rangle = E_n (S) | n, S, M \rangle \; .
\labeq{eigen}
\end{equation}
The states are classified according to their total spin quantum number $S=S_{\rm min}, ... , S_{\rm max}-1 , S_{\rm max}$ and their total magnetic quantum number $M=-S,...,S-1,S$. 
Here, $S_{\rm max}=Ls$ and $S_{\rm min}=0$ for $s$ even or $L$ even, while $S_{\rm min}=1/2$ otherwise.
Throughout of the paper, we assume an even number of sites $L$ for the sake of simplicity.
Further, $n$ labels the eigenstates in the invariant subspace with fixed $S$ and $M$.
For a given $S$, the SU(2) invariance implies the eigenenergies $E=E_n(S)$ to be $2S+1$-fold degenerate and independent of $M$.

For the present case of ferromagnetic exchange coupling $J_{ij} > 0$, the total spin quantum number is maximal, $S = S_{\rm max} = L s$, for a ground state.
It is easy to see that the fully polarized state with all spins pointing into $+z$ direction, 
$| F \rangle \equiv | m_1 = s \rangle \otimes \cdots \otimes | m_L = s \rangle$, 
is a ground state. 
Assuming all exchange interactions to be equal between nearest neighbors, $J_{ij}=J >0$, the ground-state energy is given by 
$E_0(S_{\rm max}) = - (J/2) L s^2 \overline{z}$ where $\overline{z} = \sum_{i=1}^L z_i / L$ is the average of the coordination numbers $z_i$.

The $2S_{\rm max}+1 = 2Ls+1$-fold degeneracy of the ground-state energy will partly be lifted by the anisotropy $H_{\rm ani}$, \refeq{hamani1} or \refeq{hamani2}.
$H_{\rm ani}$ breaks the SU(2) symmetry, $[\ff S^2, H_{\rm ani}] \ne 0$, but preserves the rotational symmetry with respect to the easy axis ($z$ axis): $[S_z,H_{\rm ani}]=0$.
The anisotropic eigenstates $|m, M \rangle$ are therefore characterized by $M$ and by an additional quantum number $m$ labeling the states in the subspace with fixed $M$.
The corresponding eigenenergies $E_m(M)$ will depend on $M$ with 
\begin{equation}
  E_m(M) = E_m(-M) \; ,
\end{equation}
resulting in a symmetric anisotropy barrier with a height $\Delta E$ given by the respective ground-state energies of the full Hamiltonian in the invariant subspaces with $M=0$ and $M=S_{\rm max}$:
\begin{equation}
  \Delta E = E_0(M=0) - E_0(M=S_{\rm max}) > 0 \: .
\end{equation}

\begin{figure}[t]
  \includegraphics[width=0.9\columnwidth]{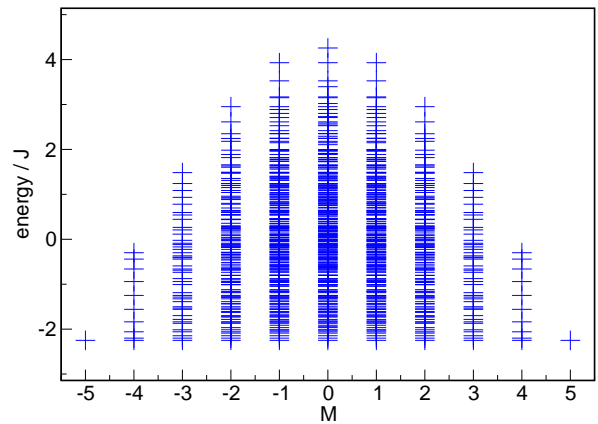}
\caption{
Energy eigenvalues of the isotropic $s=1/2$ Heisenberg model \refeq{ham} on a linear chain of $L=10$ sites and open boundaries as obtained numerically by exact diagonalization. Eigenvalues are classified according to the total magnetic quantum number $M=-S_{\rm max}, ..., S_{\rm max}$ with $S_{\rm max}=5$. The nearest-neighbor exchange coupling $J=1$ sets the energy scale.
}
\label{spectrum}
\end{figure}

\begin{figure}[b]
  \includegraphics[width=0.9\columnwidth]{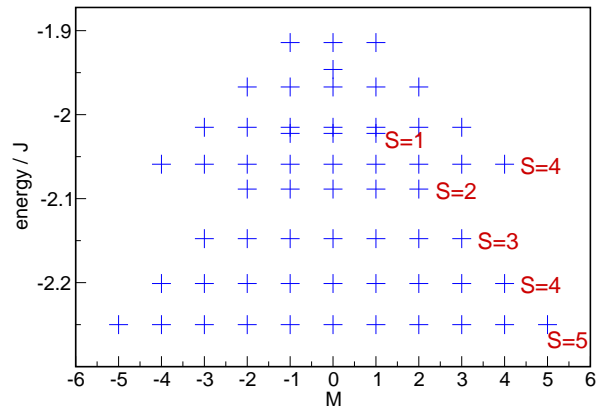}
\caption{
The same as Fig.\ \ref{spectrum} but on enlarged energy scale close to the ground-state energy. 
The total spin quantum numbers of the different spin multiplets are indicated.
}
\label{spectrum1}
\end{figure}

\begin{figure}[t]
  \includegraphics[width=0.9\columnwidth]{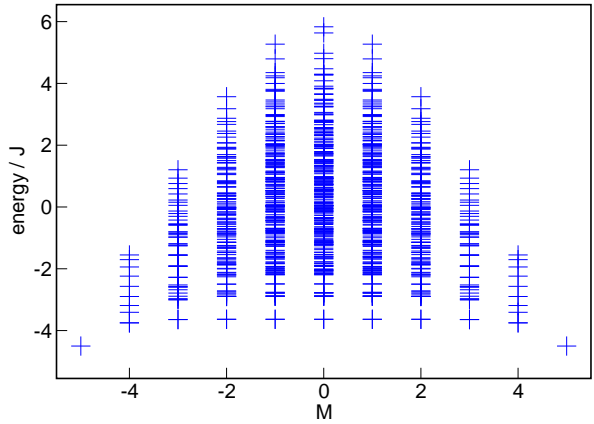}
\caption{
The same as Fig.\ \ref{spectrum} but with the coupling-anisotropy term, \refeq{hamani2}, included. 
A constant anisotropy $\Delta_{ij} = \Delta$ between nearest neighbors is assumed. 
Calculations have been performed for $\Delta = J$.
}
\label{spectrumani}
\end{figure}

For systems with a moderately large Hilbert space, i.e.\ for small $s$ and small number of lattice sites $L$, the energy eigenvalue problem can be solved numerically. 
In case of the anisotropic problem $H_J + H_{\rm ani}$, conservation of $S_z$ can be exploited only. 
The dimension of the invariant subspace $\ca H(M)$ is given by: \cite{Fel68,Sch04}
\begin{eqnarray}
  \dim \ca H(M) &=& \sum_{k=0}^{[(S_{\rm max} - M )/(2s+1)]} (-1)^k 
  \left( 
  \begin{array}{cc}
  L \\ k \\
  \end{array}
  \right)
\nonumber \\
  &\times&
  \left( 
  \begin{array}{cc}
  L-1+S_{\rm max}-M-(2s+1)k\\ L-1  \\
  \end{array}
  \right) \: 
\nonumber \\
\end{eqnarray}
where the binomial coefficients $\left( \begin{array}{cc} a \\ b \\ \end{array} \right) = 0$ for $b>a$, and where $[x]$ denotes the greatest integer with $[x]\le x$.
For $s=1/2$ full diagonalization is easily possible for systems with $L=14$ spins.

A simple example for a spin-$s=1/2$ Heisenberg chain with and without coupling anisotropy $\Delta$ on $L=10$ sites with open boundaries is given by Figs.\ (\ref{spectrum}), (\ref{spectrum1}) and (\ref{spectrumani}).
The perturbed ($\Delta >0$) ground-state energy in the one-dimensional $M=S_{\rm max}$ subspace is easily calculated as $E_0(M=S_{\rm max}) = - (J/2+\Delta/2) L s^2 \overline{z}$.
The $M=\pm(S_{\rm max}-1)$ subspaces are reached by a single spin flip, i.e.\ excitation of a magnon.
As there must be as many magnons as lattice sites, the dimension of the subspaces is given by $L$ each.
Their discrete energy spectrum can be seen in Figs.\ (\ref{spectrum}) and (\ref{spectrumani}) for $\Delta=0$ and $\Delta = J$, respectively. 

The low-energy sector of the isotropic spectrum for arbitrary $M$ is shown in Fig.\ \ref{spectrum1}: 
Ground states have the maximal total spin $S=S_{\rm max}$. 
The ground-state energy is $2S_{\rm max}+1$-fold degenerate.
Excited states with $M=\pm(S_{\rm max}-1)$, corresponding to magnon excitation, have total spin $S=S_{\rm max}-1$. 

Fig.\ \ref{spectrumani} demonstrates the effect of an anisotropy term in the Hamiltonian. 
As the energy eigenstates are superpositions of states with different $S$ (but the same $M$) in this case, their energy becomes $M$ dependent. 
Consequently, an anisotropy barrier develops which, for a very strong anisotropy $\Delta=J$, is large as compared to the finite-size gap between the ground states and the first excited states.
Still, there is a residual degeneracy of the two ground states $|F\rangle$ and $|-F\rangle$ in the subspaces $M=\pm S_{\rm max}$ which implies that thermal fluctuations destroy ferromagnetic order.
However, switching between $|F\rangle$ and $|-F\rangle$ now requires that thermal fluctuations must overcome the barrier.
The anisotropy thus leads to a superparamagnetic stabilization of the magnetic state on a certain time scale as described by the N\'eel-Brown model. \cite{Nee49,Bro63}

\section{Perturbative derivation of the macrospin model}
\label{sec:pert}

In the weak-anisotropy limit, the different microscopic spins are tightly bound together by the ferromagnetic exchange coupling and form a huge macrospin which rotates from $+z$ to $-z$ direction, i.e.\ the microspins rotate coherently.
To make this intuitive argument rigorous and quantitative, we derive the macrospin model as an effective low-energy model by means of first-order perturbation theory in $H_{\rm ani}$ in the following.
This will result in a linear dependence of $\Delta E$ on the anisotropy strength. 
Non-trivial dependencies, however, may be expected with respect to $L$, $s$ and the matrix $J_{ij}$ of exchange-coupling constants.

First-order perturbation theory requires to compute the matrix element
$\langle S_{\rm max}, M' | s_{iz} s_{jz} | S_{\rm max}, M \rangle$
where the states $|S,M\rangle$ form an orthonormal common eigenbasis of $\ff S^2$ and $S_z$.
Note that this includes the case of coupling ($i\ne j$) and single-site ($i=j$) single-site anisotropy.
To find the action of $s_{iz} s_{jz}$ onto the state $| S_{\rm max}, M \rangle$ we write 
\begin{equation}
  | S_{\rm max}, M \rangle
  =
  \frac{\sqrt{(S_{\rm max} + M)!}}{\sqrt{(S_{\rm max}-M)! (2S_{\rm max})!}}
  S_-^{S_{\rm max}-M}
  | F \rangle
\labeq{spinstate}
\end{equation}
in terms of the fully polarized state with all spins pointing in $+z$ direction, 
$| F \rangle \equiv| S_{\rm max}, M=S_{\rm max} \rangle = | m_1 = s \rangle \cdots | m_L = s \rangle$.
Here $S_- = S_x - i S_y$. 
\refeq{spinstate} is obtained by straightforward spin algebra.
Using the commutator
\begin{eqnarray}
  [s_{iz}s_{jz}, S_-^k] 
  &=&
  - k S_-^{k-1} (s_{i-}s_{jz} + s_{j-} s_{iz}) 
\nonumber \\
  &+& k \delta_{ij} S_-^{k-1} s_{i-} 
\nonumber \\
  &+& \frac{k(k-1)}{2} S_-^{k-2} (s_{i-} s_{j-} + s_{j-}s_{i-})
\end{eqnarray}
where $s_{i\pm} = s_{ix} \pm is_{iy}$,
we are left with matrix elements of the form $\langle S_{\rm max}, M' | \ca O | F \rangle$
where $\ca O$ contains microspin operators $s_{...}$ on the right-hand side acting on $|F\rangle$ as well as macrospin operators $S_{...}$ on the left-hand side which are considered to act on $\langle S_{\rm max}, M |$.
The effect of the former operators is trivial.
The effect of the latter ones is obtained with the help of the relation
\begin{eqnarray}
  S_+^k | S_{\rm max}, M \rangle
  &=&
  \frac{\sqrt{(S_{\rm max} + M + k)! (S_{\rm max} - M)!}}{\sqrt{(S_{\rm max} + M)! (S_{\rm max} - M - k)!}}
\nonumber \\
  &\times& | S_{\rm max}, M + k \rangle \: .
\end{eqnarray}
Using
\begin{equation}
  \langle S_{\rm max}, M | s_{i-} | F \rangle
  =
  \frac{2s}{\sqrt{2S_{\rm max}}}   
\end{equation}
for $M=S_{\rm max}-1$
and 
\begin{eqnarray}
  \langle S_{\rm max}, M | s_{i-} s_{j-} | F \rangle 
  &=&
  \frac{8s^2}{\sqrt{4S_{\rm max} (2S_{\rm max} -1)}}   \quad (i\ne j)
\nonumber \\
  &=&
  \frac{4s(2s-1)}{\sqrt{4S_{\rm max} (2S_{\rm max} -1)}}   \quad (i=j)
\nonumber \\
\end{eqnarray}
for $M=S_{\rm max}-2$, a straightforward calculation leads to
\begin{eqnarray}
  && \langle S_{\rm max}, M | s_{iz} s_{jz} | S_{\rm max}, M' \rangle
  \nonumber \\
  && = \delta_{MM'} \,\left[
  s^2 + (2s-\delta_{ij}) \: \frac{M^2 - L^2 s^2}{2L^2 s - L}
  \right] \: .
\labeq{corrf}
\end{eqnarray}

This is a remarkably simple result which may be used to compute the anisotropy energy barrier $E_0(M)$ in different situations.
To give an example, we consider the model $H=H_J+H_{\rm ani}$ where a constant exchange interaction $J_{ij}=J$ between nearest neighbors is assumed (and likewise for $\Delta_{ij}$ in case of a coupling anisotropy).
This implies that geometrical properties enter via the lattice topology only.
For the case of the single-site anisotropy (\ref{eq:hamani1}), $s\ne 1/2$, we get:
\begin{eqnarray}
  E_0(M) &=& - \frac{1}{2} JLs^2 \overline{z} - D \left(
  \frac{2s-1}{2Ls-1} M^2 + s^2 \frac{L(L-1)}{2Ls-1}
  \right) \nonumber \\ &+& \ca O(D^2) \: ,
\end{eqnarray}
while for the coupling anisotropy (\ref{eq:hamani2}) with non-zero and constant $\Delta_{ij}=\Delta$ between nearest neighbors only, we find:
\begin{eqnarray}
  E_0(M) &=& - \frac{1}{2} JLs^2 \overline{z} - \Delta \,\overline{z} \left(
  \frac{s}{2Ls-1} M^2 - \frac{1}{2} \frac{Ls^2}{2Ls-1}
  \right) \nonumber \\ &+& \ca O(\Delta^2) \: ,
\labeq{e0ofm}
\end{eqnarray}
where $\overline{z} = \sum_i z_i / L$ is the average coordination number.
This corresponds to a macrospin model \refeq{hammacro} with
\begin{equation}
  D_{\rm macro} = \frac{2s-1}{2Ls-1} D
\labeq{macro1}
\end{equation}
for the single-site anisotropy, and 
\begin{equation}
  D_{\rm macro} = \frac{\overline{z}s}{2Ls-1} \Delta
\labeq{macro2}
\end{equation}
in case of the coupling anisotropy. 
The anisotropy energy barrier is given by $\Delta E = D_{\rm macro} S_{\rm max}^2$.
Eqs.\ (\ref{eq:macro1}) and (\ref{eq:macro2}) provide an explicit and quantitative relation between the anisotropy parameter of the effective low-energy macrospin model and the microscopic model parameters.

Quite generally we note that this relation is non-trivial.
In the limit $L\to \infty$ we find $D_{\rm macro} \propto 1/L$ formally. 
However, for fixed $D$ or $\Delta$, the macrospin approximation becomes less accurate with increasing system size $L$:
Eventually the finite-size gap, which controls the validity of the perturbative treatment, is of the same order of magnitude as the anisotropy strength, i.e.\ the expansion parameter.
The classical limit is reached for $s \to \infty$. 
Here, $D_{\rm macro} = D/L$ or $D_{\rm macro} = (\overline{z}/2) \Delta/L$, respectively. 
The validity of the macrospin approximation for the classical model is discussed below.

\section{Limitations of the macrospin model}
\label{sec:lim}

\begin{figure}[t]
  \includegraphics[width=0.9\columnwidth]{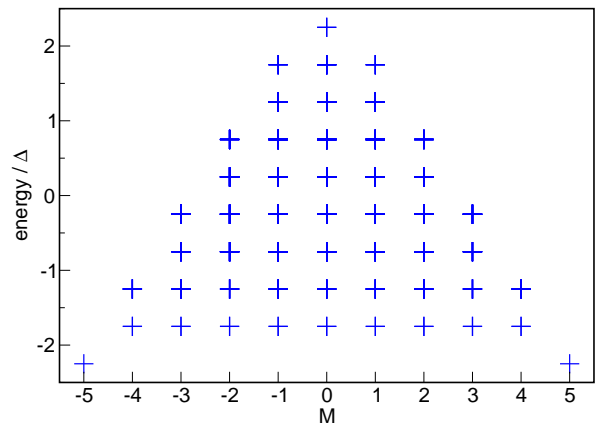}
\caption{
The same as Fig.\ \ref{spectrum} but for $J=0$ and $\Delta=1$.
}
\label{ising}
\end{figure}

It is easily seen that the limits of weak and of strong anisotropy correspond to two qualitatively different ways of how to overcome the barrier.
Consider the model system discussed above but with a large coupling anisotropy $\Delta \gg J$.
In the extreme case $J=0$, we recover the Ising model, $H=H_\Delta$, which exhibits a highly degenerate energy spectrum.
It is displayed in Fig.\ \ref{ising}. 
Still there are two degenerate ground states in the $M=\pm S_{\rm max}$ sectors which are separated by an energy barrier.
However, the barrier is entirely flat and thus qualitatively different from the quadratic trend given by the macrospin model.
The tendency towards a flat barrier is already seen in Fig.\ \ref{spectrumani} for $\Delta=J$.

The energy eigenstates are constructed trivially in the Ising limit. 
In the ground state with $M=S_{\rm max}$ the spins are aligned ferromagnetically, i.e.\ $|S_{\rm max} , S_{\rm max} \rangle = | \uparrow ,\uparrow, ... ,\uparrow \rangle$.
Its energy is $E_0(M=S_{\rm max}) = - (\Delta/2) L s^2 \overline{z} = -2.25$ for the example shown in the figure.
The ground state with $M=S_{\rm max}-1$ is found by flipping a single spin, i.e.\ $|S_{\rm max} , S_{\rm max}-1 \rangle = | \downarrow ,\uparrow, ... ,\uparrow \rangle$. 
The corresponding excitation energy is $\Delta E = 2 \Delta s^2 = 0.5$, resulting in $E_0(M=S_{\rm max}-1) = -1.75$. 
Shifting the ``domain wall'' to the right, i.e.\ $|S_{\rm max} , M \rangle = | \downarrow , ... , \downarrow, \uparrow, ... ,\uparrow \rangle$, is possible without further energy cost.
Hence, nucleation and domain-wall propagation is the apparent mechanism for magnetization reversal in the Ising limit.

\begin{figure}[t]
  \includegraphics[width=0.65\columnwidth]{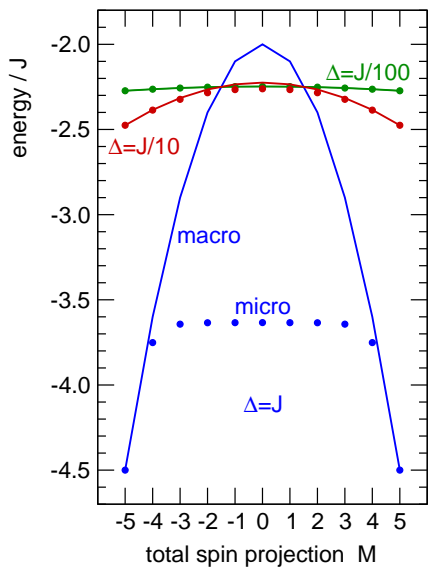}
\caption{
Anisotropy barrier $E_0(M)$ as obtained from the macrospin model (\refeq{e0ofm}, lines) compared to the exact-diagonalization result (points) for three different anisotropy strengths as indicated. 
Results for an open chain of $L=10$ spins with $s=1/2$ and nearest-neighbor exchange and coupling anisotropy.
}
\label{barrier}
\end{figure}

The breakdown of the macrospin approximation happens for still weak anisotropies but is gradual rather than abrupt.
This is demonstrated by the exact-diagonalization results for the anisotropy-energy barrier $E_0(M)$ shown in Fig.\ \ref{barrier}.
For strong anisotropy $\Delta = J$ (corresponding to Fig.\ \ref{spectrumani}), the macrospin approximation \refeq{e0ofm} completely fails and strongly overestimates the barrier height $\Delta E$.
With decreasing $\Delta$, the macrospin approximation becomes more and more reliable. 
Still, at $\Delta = J/10$ deviations from the exact-diagonalization results are visible on the scale of Fig.\ \ref{barrier}.
If interpreted within the classical theory, this is still significant as the energy barrier affects the thermal activation rate exponentially strong.

\begin{figure}[b]
  \includegraphics[width=0.75\columnwidth]{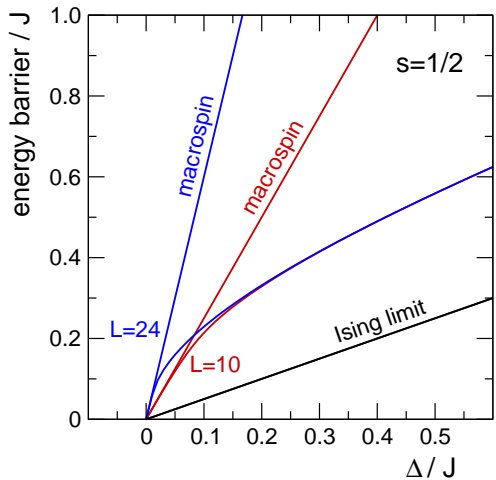}
\caption{
Height of the anisotropy energy barrier $\Delta E$ as a function of the (coupling) anisotropy strength $\Delta$ for open chains of $L=10$ and $L=24$ spins with $s=1/2$ as obtained by the macrospin model (blue lines, \refeq{macro2}) and by the full model (red lines). Black line: $\Delta E$ as a function of $\Delta$ for the Ising limit with $J=0$ for both, $L=10$ and $L=24$.
}
\label{size}
\end{figure}

The fact that the breakdown of the macrospin model with increasing strength of the anisotropy is gradual rather than abrupt also implies a gradual change between the corresponding reversal mechanisms.
Within a quantum model, there are no well-defined phase boundaries separating coherent rotation from nucleation and domain-wall propagation.
This becomes obvious again in Fig.\ \ref{size} where the anisotropy energy barrier $\Delta E$ is shown as a function of $\Delta$.
The macrospin and the exact-diagonalization results start to deviate from each other at arbitrarily weak $\Delta$, i.e.\ already at second order in the anisotropy strength. 

Nevertheless, a typical anisotropy strength can be identified that marks the smooth crossover from coherent rotation to nucleation or the qualitative breakdown of the macrospin approximation.
Fig.\ \ref{size} shows that this crucially depends on the system size $L$. 
For $L=10$ spins (red lines), the macrospin model is valid up to much stronger $\Delta$ as compared to the case with $L=24$ spins (blue lines).
Generally, with increasing $L$, substantial deviations from the linear trend of $\Delta E (\Delta)$ are found for weaker and weaker $\Delta$.
In the strong-anisotropy limit, $\Delta E (\Delta)$ approaches a linear trend again which is given by the Ising limit (black line). 
Note that for the one-dimensional model considered here, the domain-wall energy and thus $\Delta$ in the strong-anisotropy limit is independent of the system size while in the macrospin or weak-anisotropy limit $\Delta E$ is roughly proportional to $L$ (see \refeq{e0ofm}).

Calculations for $L=24$ spins with $s=1/2$ can no longer be carried out by full diagonalization. 
The present results have been obtained by employing the Lanczos algorithm. \cite{Lan50,LG93}
This approximates the ground-state energy $E_0(M)$ of $H$ in the invariant subspace $\ca H(M)$ by the ground-state energy of $H$ in a Krylov space 
\begin{equation}
  \ca K_n(M) = \mbox{span} \{|i,M\rangle, H |i,M\rangle,..., H^{n-1} |i,M\rangle\} \subset \ca H(M)
\end{equation}
of dimension $n\ll \dim \ca H(M)$ that is spanned by the states $|i,M\rangle, H |i,M\rangle,..., H^{n-1} |i,M\rangle$ where $|i,M\rangle \in \ca H(M)$ is an arbitrary initial state.
Typically, $n \approx 100$ Krylov-space dimensions are fully sufficient to obtain an excellent approximation which, on the scale of the figures, cannot be distinguished from the exact result.

\begin{figure}[t]
  \includegraphics[width=0.85\columnwidth]{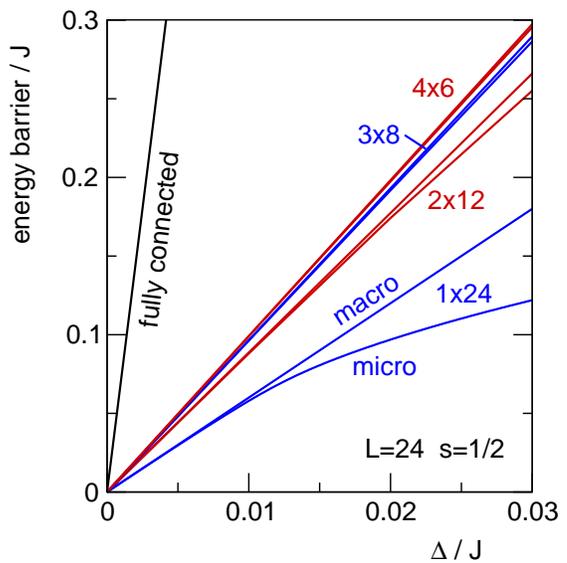}
\caption{
Height of the anisotropy energy barrier $\Delta E$ as a function of the (coupling) anisotropy strength $\Delta$ for systems of $L=24$ spins with $s=1/2$ as obtained by the macrospin model (\refeq{macro2}) and by the full model ($\Delta E$ in the full model is always smaller). Results for different planar geometries as indicated. $\Delta E$ for the fully connected model, where the macrospin approximation is exact, is shown for comparison (black line).
}
\label{coord}
\end{figure}

The extent of applicability of the macrospin model strongly depends on the system's geometry. 
It turns out to become more and more reliable with increasing average coordination number $\overline{z}$.
This is demonstrated with Fig.\ \ref{coord} where again macrospin and exact-diagonalization (Lanczos) results for the $\Delta$ dependence of the energy barrier are compared. 
Here, the system size is kept fixed to $L=24$ while the system geometry varies from a one-dimensional chain to a compact two-dimensional array, i.e.\ the average coordination number $\overline{z}$ increases.
The corresponding increasingly better agreement of the macrospin approximation with the exact-diagonalization results is easily understood by consideration of the extreme case (see black line in Fig.\ \ref{coord}): 
For $z_i=L-1=\overline{z}$, i.e. for the fully connected model with $J_{ij}=J$ and $\Delta_{ij}=\Delta$ for arbitrary pairs $i,j$, the Hamiltonian reduces to
\begin{equation}
H=-J \ff S^2 - \Delta \, S_z^2 + \mbox{const}
\end{equation}
and thus
\begin{equation}
E_0(M) = - \Delta \, M^2 + \mbox{const} \; ,
\end{equation}
i.e.\ in this limit the macrospin approximation is trivially exact.

\section{Critical anisotropy strength in the classical limit}
\label{sec:class}

\begin{figure}[t]
  \includegraphics[width=0.85\columnwidth]{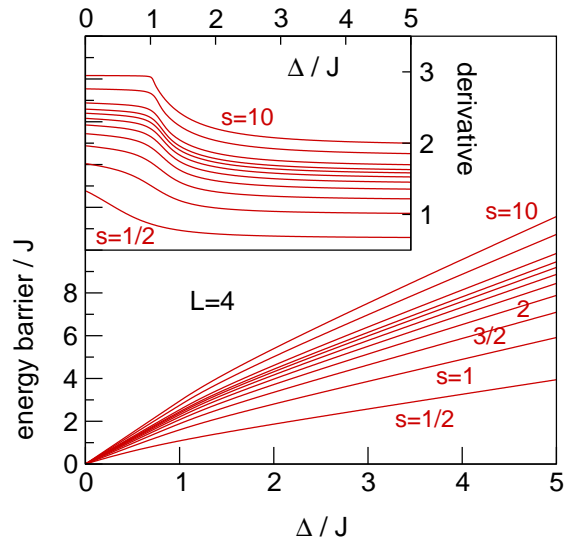}
\caption{
$\Delta$ dependence of energy barrier $\Delta E$ for open chains with $L=4$ spins and different spin quantum numbers $s$ as indicated. Exact results for the full model $H_J + H_\Delta$ after rescaling $\ff s_i \to \ff s_i / \sqrt{s(s+1)}$ for 
$s=1/2,1,3/2,2,5/2,3,7/2,4,5,10$. The inset shows the first derivative of $\Delta E$ with respect to $\Delta$. 
}
\label{class}
\end{figure}

The limit of the classical anisotropic Heisenberg model should be recovered for spin quantum number $s\to \infty$. 
Fig.\ \ref{class} shows the anisotropy barrier $\Delta E$ as a function of the coupling anisotropy strength $\Delta$ as obtained from the Lanczos technique for a small chain of $L=4$ sites.
This permits a study of the spin model with quantum numbers up to $s=10$.
To compare the results for different $s$, we consider rescaled parameters, 
\begin{eqnarray}
  J_{ij} &\mapsto& J_{ij} / s(s+1) \; ,
\nonumber \\
  \Delta_{ij} &\mapsto& \Delta_{ij} / s(s+1) \; ,
\end{eqnarray}
which corresponds to a rescaling $\ff s_i \mapsto \ff s_i / \sqrt{s(s+1)}$ of the spin variables.
Disregarding quantum fluctuations, this amounts to a classical spin of unit length $|\ff s_i|=1$ in the limit $s\to \infty$.

For $s=1/2$ we again note the smooth crossover from the macrospin limit and coherent rotation at weak $\Delta$ to the Ising limit and nucleation and domain-wall propagation at strong $\Delta$. 
In the quantum model the breakdown of the macrospin approximation starts immediately, i.e.\ the linear-in-$\Delta$ trend of the barrier is hardly visible as is apparent from the inset of Fig.\ \ref{class}. 
The crossover to the Ising limit takes place quickly.

For $s=1$, the range of anisotropy strengths where an almost linear trend is found, increases while the crossover to the Ising limit is seen to be delayed. 
With increasing spin quantum number $s$, these changes continue in a systematic way and become more and more pronounced.

From the result for $s=10$, one can easily anticipate the $s\to \infty$ limit: 
Here the anisotropy barrier is strictly linear in $\Delta$ for weak anisotropies up to a certain critical value $\Delta_c / J=1$.
At $\Delta_c$ we find a cusp, i.e.\ a discontinuous jump of $\partial^2 \Delta E / \partial \Delta^2$.
Beyond this point, a fairly broad crossover region follows until finally a linear-in-$\Delta$ trend is established again in the $\Delta \to \infty$ limit.

The strong-anisotropy limit is easily understood, as quantum fluctuations are suppressed anyway for all $s$. 
For weak anisotropy, on the other hand, the macrospin approximation is found to be exact in a finite range up to $\Delta = \Delta_c$, i.e.\ perturbative corrections of second and higher order in $\Delta$ are decreasing with increasing $s$ and finally vanish exactly in the classical limit $s\to \infty$. 
For all $\Delta < \Delta_c$, magnetization reversal takes place by a coherent rotation of completely correlated spins in the classical case. 
For any finite $s$, on the other hand, quantum fluctuations break up this perfect correlation.
The magnetic moments $\langle \ff s_i \rangle$ are still aligned to the $z$ axis during the reversal -- we have $\langle \ff s_i \rangle \propto \ff e_z$ since $S_z$ is a conserved quantity -- but the effect of fluctuations can be seen in the spin-spin correlation functions:
In the $M=0$ sector, for example, we find $\langle s_{iz} s_{jz} \rangle = 0$ classically ($s\to \infty$) and 
$\langle s_{iz} s_{jz} \rangle = s^2 (1-1/(1-1/2Ls))$ in the quantum case for sufficiently small $\Delta$ (see \refeq{corrf}), i.e.\ a small quantity of the order $\ca O(L^{-1})$ which is independent of $i$ and $j$. 
With increasing $\Delta$ at finite $s$, however, $\langle s_{iz} s_{jz} \rangle$ gradually becomes more and more site dependent.
As a function of $j$ and fixing $i$ to one of the chain edges, $\langle s_{iz} s_{jz} \rangle$ is monotonically decreasing in the $M=0$ sector.
Spins at the chain edges are oriented antiferromagnetically: $\langle s_{iz} s_{jz} \rangle < 0$ for $i=1$ and $j=L$.
Hence, for any finite $\Delta$ in the quantum case there are already indications for domain-wall formation while in the classical limit there is a clear-cut distinction, given by a critical coupling $\Delta_c$, between the two reversal mechanisms.

To understand the origin of the ``phase transition'' in the classical limit and to find a means to compute $\Delta_c$ for arbitrary system size and geometry, we first performed numerical minimizations of the total energy $E(\{ \ff s_1, ..., \ff s_L \})$ of the classical model $H = H_J + H_{\rm ani}$ for both, $H_{\rm ani} = H_\Delta$ and $H_{\rm ani}=H_D$.
Arbitrary spin configurations $\{ \ff s_1, ..., \ff s_L \}$ with $|\ff s_i|=1$ and satisfying the constraint $\sum_{i=1}^L s_{iz} = M$ are considered.
For $M=0$ this provides the classical barrier height and the corresponding optimal spin configuration.
Calculations have at first been done for different small system sizes $L$, typically $L=4$, to reproduce the critical point anticipated from the results of Fig.\ \ref{class}.

\begin{figure}[t]
  \includegraphics[width=0.99\columnwidth]{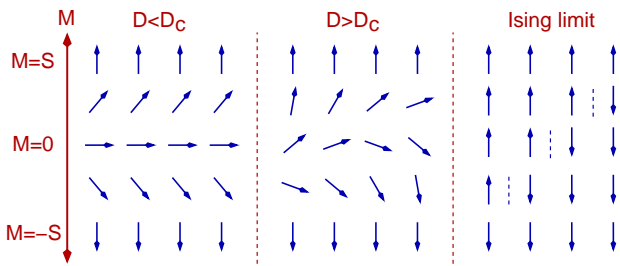}
\caption{
Sketch of the different magnetization-reversal mechanisms in the classical model for anisotropy strength $D$ smaller (macrospin model, coherent rotation) or larger (nucleation) than the critical value $D_c$ and for infinite $D$ (nucleation, vanishing domain-wall width) as verified by calculations for $L=4$ spins. 
}
\label{mechan}
\end{figure}

We found the optimal spin configuration with minimal total energy for a given $M$ to be coplanar in all cases, i.e.\ there is a coordinate frame with $s_{iy}=0$ for all spins.
As dictated by the symmetry of the model, configurations that transform into each other by rotations around the $z$ axis are degenerate.
For a chain geometry, there is also degeneracy between configurations obtained by mirror transformations, $\ff s_i \to \ff s_{L+1-i}$, and nucleation can thus start from both edges with equal probability.
Furthermore, for $M=0$ we have $s_{i,z} = - s_{L-i+1,z}$ in the optimal configuration.
The results can be summarized by the sketch of the optimal spin configurations at a given $M$ in Fig.\ \ref{mechan}.
Results for coupling and for single-site anisotropy are found to be qualitatively the same.
As is seen from Fig.\ \ref{mechan}, the different mechanisms for magnetization reversal in the two limits of weak and strong anisotropy are reproduced from the calculations for small $L$.

More important, however, the physical origin of the critical point can be identified easily:
At $\Delta = \Delta_c$ (or at $D=D_c$, respectively), and approaching the critical point from the weak-anisotropy side, the optimal spin configuration exhibits an instability towards a non-collinear ordering.
As sketched in Fig.\ \ref{mechan}, the instability develops simultaneously for any $M$.

This observation can be exploited to calculate the critical anisotropy strength for arbitrary system size and system geometry in the following way:
Using $s_{iy}=0$, we parameterize the spin variables as $\ff s_i=(\sqrt{1-m_i^2},0,m_i)$ with $-1 \le m_i \le 1$. 
This yields the energy functional in the form
\begin{eqnarray}
  && E(\{m_1,...,m_L\}) 
  =
  - \frac{1}{2} \sum_{ij} D_{ij} m_i m_j
\nonumber \\
  && - \frac{1}{2} \sum_{ij} J_{ij} \left(
  \sqrt{1-m_i^2}\sqrt{1-m_j^2} + m_i m_j
  \right) \: ,
\end{eqnarray}
which comprises both, the coupling and the single-site anisotropy case.
To find the critical point, it is sufficient to concentrate on $M=0$ and to compute
the magnetic response of the system in the $M=0$ state with spin configuration $m_1= \cdots = m_L=0$. 
Close to the critical point, we can assume $m_i \ll 1$. 
Expanding the total energy in powers of $m_i$ then yields
\begin{equation}
  E(\{m_1,...,m_L\}) 
  =
  \mbox{const.} + \frac{1}{2} \sum_{ij} \chi^{-1}_{ij} m_i m_j + {\cal O}({m_i^4})
\end{equation}
with the inverse susceptibility matrix
\begin{equation}
  \chi^{-1}_{ij}
  =
  - J_{ij} + J_i \delta_{ij} - D_{ij} \: ,
\end{equation}
and where we have defined $J_i \equiv \sum_j J_{ij}$. 
We assume the model parameters as real and symmetric, $J_{ij}= J_{ji}$ and $D_{ij}=D_{ji}$. 

Note that approaching the isotropic limit with anisotropy strength $D\to 0$, the susceptibility matrix exhibits an eigenvalue approaching zero.
This corresponds to an instability of the state towards ferromagnetic ordering with all $m_i > 0$ and indicates the incipient degeneracy of states with different $M$.

For finite $D$, this eigenvalue is negative and thus implies an indefinite susceptibility.
This indicates that, by construction, $\chi$ refers to the thermodynamically unstable excited state with $M=0$ and $m_i=0$. 
The {\em local} instability of this state towards a non-collinear alignment of the magnetic moments at the critical anisotropy strength $D_c$ shows up as a zero of another eigenvalue of $\chi$.
For a short chain of $L=4$ spins with coupling anisotropy this happens exactly at $\Delta_c = J$, in agreement with the Lanczos results displayed in Fig.\ \ref{class}. 
The eigenvector corresponding to the zero eigenvalue describes the critical spin profile which agrees with the $M=0$ spin configuration for $D>D_c$ sketched in Fig.\ \ref{mechan}.

\begin{figure}[t]
  \includegraphics[width=0.99\columnwidth]{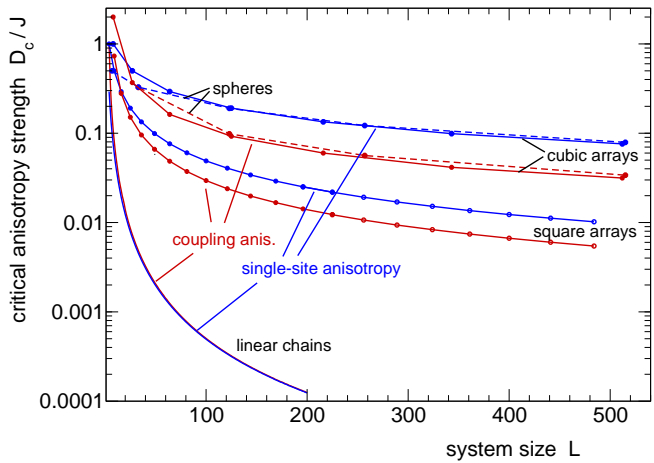}
\caption{
System-size dependence of the critical value for the anisotropy strength $D_c$ in units of $J$ for one-dimensional open chains, for two-dimensional square arrays and for three-dimensional simple cubic arrays and spheres with open boundaries. Blue lines: single-site anisotropy. Red lines: coupling anisotropy. 
}
\label{critical}
\end{figure}

Results for the critical anisotropy strength as a function of the system size are shown in Fig.\ \ref{critical} for both, coupling and single-site anisotropy and for systems of different dimensions $d$. 
In all cases we considered systems with open boundaries.
Generally, the anisotropy strength up to which the macrospin approximation is found to be valid, decreases rapidly with system size but depends only weakly on the type of the anisotropy term. 
There is also a weak dependence on the form of the particle as is obvious from the comparison of cubic arrays and spheres in $d=3$ dimensions.

For small nanosystems, the dimension and shape dependence of $D_c$ is somewhat irregular.
For a larger number of classical spins, on the other hand, the critical anisotropy strength for which the macrospin model can be applied, is the stronger the higher the dimension of the nanostructure.
In fact, in the limit $L\to \infty$, the different results are found to follow a simple power law,
\begin{equation}
  D_c(L) \propto L^{-2/d} = l^{-2} \; ,
\label{eq:scal}
\end{equation}
where $d$ is the dimensionality, $L$ the number of spins and $l = L^{1/d}$ the {\em linear} extension of the system.

This can easily be explained by comparing the energy barrier for coherent rotation $\Delta E_{\rm cor.rot.}$ with the barrier for nucleation and domain-wall propagation $\Delta E_{nucl.}$. 
While the volume of the system $\propto l^d$ is relevant for $\Delta E_{\rm cor.rot.} \propto D \, l^d$, the $d-1$-dimensional domain wall of width $\sigma(D) \propto \sqrt{J/D}$ \cite{Kit49,Mid63} essentially determines $\Delta E \propto D \, \sigma(D) \, l^{d-1}$.
The critical anisotropy is then obtained from $D_c \, l^d \propto D_c \sqrt{J/D_c} \, l^{d-1}$, resulting in the scaling law (\ref{eq:scal}).

\section{Conclusions}
\label{sec:con}

Anisotropy terms coupled to the ferromagnetic Heisenberg model of many-spin nanosystems give rise to an energy barrier $\Delta E$ that crucially determines the thermal activation of magnetization reversal and thus the magnetic stability of nanosystems.
Employing perturbation theory, exact diagonalization and Lanczos as well as classical total-energy minimization and stability analysis, different quantum effects have been revealed that substantially affect the barrier height.

Quantum effects are absent in the limit of strong anisotropy $D$. 
For $D/J \to \infty$ the Ising model is approached, i.e.\ a classical model where there are no quantum fluctuations. 
The magnetization-reversal mechanism in the Ising limit is given by nucleation at the edge or surface of the system followed by the propagation of a domain wall with minimal (unit) width.
With decreasing anisotropy strength, the domain-wall width increases.
Finally, a critical value $D = D_c$ is reached where the reversal mechanism changes to a coherent rotation of the spins that are tightly aligned by the exchange coupling $J$ and form a huge macrospin.
This critical anisotropy strength is well defined in the classical limit $s \to \infty$ (with model parameters rescaled by the factor $s(s+1)$) and can be computed for arbitrary dimension, system size and shape by an analysis of the local stability of the susceptibility matrix in the globally (thermodynamically) unstable excited state at the top of the barrier.
For $D<D_c$, i.e.\ for coherent rotation, the classical anisotropy-barrier height is simply given by
$\Delta E_{\rm class} = E(m_1= \cdots m_L = 0) - E(m_1= \cdots = m_L =1)$, i.e.:
\begin{equation}
  \Delta E_{\rm class} = L \, D  \; .
\label{eq:deltaeclass}
\end{equation}
This is the same as the barrier of a classical macrospin of length $\ff S=L\ff s$, composed of microscopic spins of unit length $|\ff s|=1$, as described by a macrospin model $H=D_{\rm macro} S_z^2$ with a barrier $\Delta E = D_{\rm macro} L^2$ if $D_{\rm macro} = D / L$.
Classically, and in the regime where coherent rotations takes place, the macrospin approximation is exact.

The first quantum effect manifests itself in the absence of the a well-defined critical anisotropy strength. 
As can be seen in the spin-spin correlation function $\langle s_{iz} s_{jz} \rangle$, there is rather a smooth crossover from nucleation and domain-wall propagation to coherent rotation with decreasing $D$.
This crossover becomes less and less sharp with decreasing $s$, i.e.\ as the importance of quantum fluctuations increases.

Second, for any finite anisotropy strength, the macrospin approximation is no longer correct in the quantum case.
Only within perturbation theory to first order in $D$, the macrospin model emerges as the effective low-energy theory of the anisotropic quantum Heisenberg model.
Comparing the true anisotropy-energy barrier $\Delta E$ of the many-spin system with the one of the corresponding effective macrospin model, we find
\begin{equation}
  \Delta E < \Delta E_{\rm macro} \: .
\end{equation}
In the quantum case, there is virtually never a reversal by pure coherent rotation but always some ``admixture'' of nucleation and domain-wall propagation.
The latter leads to the decrease of $\Delta E$ as compared to $\Delta E_{\rm macro}$.
This quantum effect becomes more and more pronounced with increasing anisotropy strength, with increasing system size, with decreasing dimensionality and with decreasing average coordination number.

Third, the quantum derivation of the macrospin model by means of first-order perturbation theory provides an explicit expression for the anisotropy parameter [see Eqs. (\ref{eq:macro1}) and (\ref{eq:macro2})]. 
This leads to an energy barrier that is lower than the classical one:
\begin{equation}
  \Delta E_{\rm macro} < \Delta E_{\rm class} \: .
\end{equation}
Consider a single-site anisotropy case as an example: 
After the rescaling $D \to D/s(s+1)$ we have
$\Delta E_{\rm macro} = [(2s-1)/(2Ls-1)] [L^2s^2 / s(s+1) ] D$.
Therewith, the classical result (\ref{eq:deltaeclass}) is recovered in the high-spin limit: 
$\lim_{s\to\infty} \Delta E_{\rm macro} = \Delta E_{\rm class}$. 
In the low-spin case, e.g.\ for $s=1$, which is the extreme case for single-site anisotropy, however, we have $\Delta E_{\rm macro} = (L/4) D$.
This is smaller by a factor of 4 than the classical barrier $\Delta E_{\rm class}$. 
Even in the limit of large systems $L\to \infty$, there is substantial difference,
\begin{equation}
  \Delta E_{\rm macro} = \frac{1}{2} \frac{2s-1}{s+1} L D < L D = \Delta E_{\rm class} \: ,
\end{equation}
compared to the classical limit which is approached rather slowly, i.e.\ 
$\Delta E_{\rm macro} / \Delta E_{\rm class} = 1 - 3s / s + \ca O(s^{-2})$.
We conclude that, in addition to quantum many-spin effects, there is a quantum correction of the macro-spin model itself which, for low spin quantum numbers, can be as important as the former one.

\acknowledgments
We would like to thank E.\ Y.\ Vedmedenko for helpful discussions.
The work is supported by the Deutsche Forschungsgemeinschaft within the Sonderforschungsbereich 668 (project B3).

\end{document}